\documentclass[a4paper]{article}

\usepackage{graphics} 
\graphicspath{{graphics/}} 
\usepackage{amsmath} 
\usepackage{amssymb}  

\usepackage{mathtools}

\usepackage{amsthm}
\usepackage{microtype}

\usepackage{derivative}

\newtheorem{thm}{Theorem}

\newtheorem{cor}[thm]{Corollary}

\title{\bf Trajectory tracking model-following control using Lyapunov redesign with output time-derivatives to compensate unmatched uncertainties
}

\author{
	Niclas Tietze\thanks{Control Engineering Group,  Technische Universt\"at Ilmenau, P.O.~Box 10 05 65, D-98684 Ilmenau, Germany.} \and
	Kai Wulff$^*$\thanks{Corresponding author: \texttt{kai.wulff@tu-ilmenau.de}} \and 
	Johann Reger$^*$}

\begin{document}

\maketitle
\thispagestyle{empty}
\pagestyle{empty}

\begin{abstract}
	We study trajectory tracking for flat nonlinear systems with unmatched  uncertainties using the model-following control (MFC) architecture.
	We apply state feedback linearisation control for the process and propose a simplified implementation of the model control loop which results in a simple model in Brunovsk\'{y}-form that represents the nominal feedback linearised dynamics of the nonlinear process.
	To compensate possibly unmatched model uncertainties, we employ Lyapunov redesign with numeric derivatives of the output.
	It turns out that for a special initialisation of the model, the MFC reduces to a single-loop control design.
	We illustrate our results by a numerical example.
\end{abstract}
\section{Introduction}
The model-following control scheme (MFC) is a well-known two-degree-of-freedom architecture for robust control, see Figure \ref{fig:MFC_blockdiagram_conventional}.
It consists of two control loops, namely, the model control loop (MCL) and the process control loop (PCL).
It is widely demonstrated that with a precise model of the process dynamics, the strength of the MFC lies in compensating matched perturbations, see e.g. \cite{Pre1972} and \cite{CheC1998}.
In our preliminary studies \cite{WilWR2021}, \cite{WilWR2022}, \cite{TieWR2024CDCSMC}, \cite{TieWR2024a} and \cite{TieWR2024} of the MFC architecture, we consider a nonlinear model of the nominal dynamics of the process, how that typical robust control techniques, such as sliding mode control, high-gain control \cite{KhaS1982}, \cite{YouKU1977} or Lyapunov redesign \cite{Lei1978}, \cite{Gut1979}, \cite{GutP1982}, can be applied within the framework of state feedback linearisation for both, stabilisation and trajectory tracking for minimumphase systems with matched model uncertainties.
However, the assumption of matched model uncertainties, i.e. the assumption that the model uncertainty appears in the input channel only, is restrictive.

A variety of techniques can be applied to robustify feedback linearisation with respect to unmatched uncertainties. 
E.g. in \cite{ChoW1995} high-gain feedback is applied to attenuate the effect of unmatched model uncertainties, and for minimumphase nonlinear systems with (possibly unmatched) parametric uncertainties the authors of \cite{SasK1988} and \cite{Azi2017} propose a parameter adaptive control design and an approximate feedback linearisation Lyapunov redesign, see \cite{GuaS2001}, respectively.  
In the context of sliding mode control design however, it is well-known that unmatched model uncertainties can be compensated using discontinuous feedback of the time-derivatives of the output.
In our preliminary study \cite{WulPR2020}, we stabilise a minimumphase system with unmatched perturbation using feedback linearisation and sliding mode control with a sliding surface, which depends on the time-derivatives of the output of the process, and in \cite{PosWR2020} and \cite{PosWR2023}, the influence of perturbations, that alter the relative degrees is studied.

The aim of this study is to combine the results of our preliminary studies \cite{WulPR2020} and \cite{TieWR2024a}.
We consider output tracking for a flat nonlinear system with unmatched model uncertainty, applying state feedback linearisation to the process.
It turns out that we only need to the simulation of the feedback linearised model in Brunovsk\'{y}-form, i.e. we do not require the simulation of the nonlinear model including its feedback-linearising control law.
To achieves asymptotic trajectory tracking in the presence of unmatched model uncertainties, which cannot be compensated using feedback of the system state, we apply the idea of feedback of time-derivatives of the output in \cite{WulPR2020} to the Lyapunov redesign MFC of \cite{TieWR2024a}, which was inspired by the pioneering publication \cite{SugO1993}.
Using a state transformation given by the derivatives of the flat output, we propose a Lyapunov redesign control law which compensates the unmatched uncertainty assuming bounding conditions on the model uncertainty. 

The paper is organised as follows.
Section \ref{sec:problem_definition} gives a formal problem definition.
In Section~\ref{sec:unmatched_perturbation} we demonstrate the restrictiveness of the assumption of matched model uncertainties by showing that unmatched uncertainties are invariant w.r.t a nonlinear change of coordinates.
In Section~\ref{eq:mfc_with_perturbation} the MFC architecture is introduced.
It is shown that the MFC design of our preliminary studies cannot be applied for output tracking in presence of unmatched uncertainties.
Section~\ref{sec:mfc_design} presents our main results, namely the MFC design, which considers a linear model of the feedback linearised process and achieves asymptotic output tracking in presence of unmatched model uncertainties.
For a special initialisation of the model, the design is equivalent to a single-loop control design.
In Section~\ref{sec:example} we illustrate the Lyapunov redesign MFC approach by an illustrative example.
We end with the conclusions in Section~\ref{sec:conclusion}.

\section{Problem definition}\label{sec:problem_definition}
We consider the flat nonlinear system 
\begin{subequations}\label{eq:system}
	\begin{align}
		\dot{x} &= f(x) + g(x) \, u + \Delta (x),
		\\
		y &= h(x),
	\end{align}
\end{subequations}
where $x(t)\in \mathbb{R}^n$ denotes the state, $u(t)\in \mathbb{R}$ is the control input, and $y(t)\in \mathbb{R}$ is the flat output.
The known vector fields $f:\mathbb{R}^n \mapsto\mathbb{R}^n$ and $g: \mathbb{R}^n\mapsto \mathbb{R}$ are sufficiency smooth, where $f(0) = 0$ and $g(x)\neq0$ for all $x\in \mathbb{R}^n$, and the output function $h:\mathbb{R}^n\mapsto  \mathbb{R}$ is uniformly continuous.
The function $\Delta :\mathbb{R}^n\mapsto\mathbb{R}^n$ represents an unknown, possibly unmatched model uncertainty which is sufficiently smooth.
We require the output $y=h(x)$ of the nominal system, i.e. $\Delta(x) \equiv0$, to have relative degree $r = n$ w.r.t. to the input $u$, i.e.
\begin{subequations}\label{eq:relative_degree_condition}
	\begin{align}
		\mathcal{L}_{g}\mathcal{L}_{f}^{k}h(x) &=  0 \quad \text{for } k=0,1,...,n-2, 
		\label{eq:relative_degree_condition_zero_before_input_channel}
		\\
		\mathcal{L}_{g}\mathcal{L}_{f}^{n-1}h(x)&\neq 0,
		\label{eq:relative_degree_condition_not_zero_in_input_channel}
	\end{align}
\end{subequations}
for all $x\in \mathbb{R}^n$, where $\mathcal{L}$ denotes the Lie derivative.
Similar to \cite{WulPR2020}, \cite{PosWR2020} and \cite{PosWR2023}, we further require that the relative degree $r$ is uniform w.r.t. the model uncertainty $\Delta(\cdot)$, i.e.
\begin{subequations}\label{eq:relative_degree_preserving_perturbation}
	\begin{align}
		\mathcal{L}_{g}\mathcal{L}_{f+\Delta}^{k}h(x) &= 0 \quad \text{ for } k=0,1,...,n-2,
		\\
		\mathcal{L}_{g}\mathcal{L}_{f+\Delta}^{n-1} h(x) &\neq0,
	\end{align}
\end{subequations}
where $\mathcal{L}_{f+\Delta}^k$ is the sum of all possible compositions of $\mathcal{L}_f$ and $\mathcal{L}_{\Delta}$ of length $k$.
The mappings 
\begin{align}
	\tau(x) &= \begin{bmatrix}
		h(x) & \mathcal{L}_{f}h(x) & ... & \mathcal{L}_{f}^{n-1}h(x)		
	\end{bmatrix}^\top,
	\label{eq:state_transformation}
	\\
	\tau_\mathrm{n}(x) &= \begin{bmatrix}
		h(x) & \! \mathcal{L}_{f + \Delta}h(x) & \! \! ... \! \! & \mathcal{L}_{f + \Delta}^{n-1}h(x)
	\end{bmatrix}^\top
	\label{eq:state_transformation_derivatives}
\end{align}
are assumed to be global diffeomorphisms, where a condition for this assumption to hold can be found e.g. in \cite[Section~9.1]{Isi1995}, see also \cite{SasK1988} and \cite{ByrI1991}.
Further we assume that both, the state $x$ of the process and the first $n-1$ time-derivatives of the output $y=h(x)$, i.e.
\begin{align}\label{eq:output_derivatives}
	y^{(k)} 
	= h^{(k)}(x) 
	= \mathcal{L}_{f + \Delta}^{k}h(x),
	\quad k = 0,1,...,n-1
\end{align}
are available at run-time.
The goal is to design a controller such that the output $y(t) = h(x(t))$ tracks the bounded reference signal $y_\mathrm{d}(t)$, whose time-derivatives $y_\mathrm{d}^{(i)},i=1,...,n$ are bounded and available at run-time.
The idea of applying feedback of the time-derivatives \eqref{eq:output_derivatives} of the output is to compensate, the possibly unmatched uncertainty $\Delta(\cdot)$ using the information of $\Delta$ contained in $y^{(k)} = \mathcal{L}_{f + \Delta}^{k}h(x)$. 
The time-derivatives may be obtained by sliding-mode differentiation, see e.g. \cite{Lev2009} and \cite{LevLY2017}.


\section{Unmatched uncertainties}\label{sec:unmatched_perturbation}
In this section, we show that unmatched model uncertainties pose a fundamental problem for state feedback linearisation control design.
Consider the state transformation $\xi = \tau(x)$ with $\tau(\cdot)$ in \eqref{eq:state_transformation}.
The dynamics \eqref{eq:system} of the process read
\begin{align}\label{eq:dynamics_system_transformed_coordinates}
	\!
	\dot{\xi} \!=\! A \, \xi \!+\! B  \big(
	\mathcal{L}_{f}^n h(x) \!+\! \mathcal{L}_{g}\mathcal{L}_{f}^{n-1}h(x)  u \!+\! \phi_\mathrm{m}(x)
	\big) \! + \! \phi_\mathrm{u}(x),
\end{align}
with output $y = \xi_1$, where the pair $(A,B)$ is in Brunovsk\'{y}-form, and $\phi_\mathrm{m}(\cdot)$ and $\phi_\mathrm{u}(\cdot)$ represent the matched and unmatched uncertainties in transformed coordinates, respectively,
\begin{subequations}\label{eq:pertrubation_transformed}
	\begin{align}
		\hspace{-1.5ex} \phi_\mathrm{m}(x) &= \mathcal{L}_{\Delta} \mathcal{L}_{f}^{n-1}h(x)
		\label{eq:pertrubation_transformed_matched}
		\\
		\hspace{-1.5ex} \phi_\mathrm{u}(x) &= \begin{bmatrix}
			\mathcal{L}_{\Delta}h(x) \ \mathcal{L}_{\Delta} \mathcal{L}_{f}h(x) \ ... \ \mathcal{L}_{\Delta} \mathcal{L}_{f}^{n-2}h(x) \ 0
		\end{bmatrix}^\top\! \! \! .
		\label{eq:pertrubation_transformed_unmatched}
	\end{align}
\end{subequations}
For brevity of the expression, we write $x$ as argument of the Lie derivatives in the dynamics \eqref{eq:dynamics_system_transformed_coordinates} of the transformed coordinates $\xi$ and in the uncertainty \eqref{eq:pertrubation_transformed}, although the expressions should be read for $x = \tau^{-1}(\xi)$.

The unmatched uncertainty $\phi_{\mathrm{u}}(\cdot)$ in transformed coordinates~$\xi$ vanishes if and only if the uncertainty $\Delta$ of the dynamics \eqref{eq:system} in original coordinates $x$ satisfies
\begin{align}\label{eq:model_uncertainty_without_unmatched_component}
	\mathcal{L}_{\Delta}\mathcal{L}_{f}^{k}h(x) = 0 \quad \text{for } k=0,1,...,n-2.
\end{align}
Following \cite{CasF2006} and \cite{RubEC2011}, we decompose the uncertainty $\Delta(\cdot)$ into a matched and an unmatched uncertainty $\Delta_\mathrm{m}(\cdot)$ and $\Delta_\mathrm{u}(\cdot)$, respectively.
Let
\begin{subequations}\label{eq:perturbation_decomposition}
	\begin{align}
		\Delta (x) &= \Delta_\mathrm{m}(x) + \Delta_\mathrm{u}(x),
		\label{eq:perturbation_decomposition_sum}
		\\
		\Delta_\mathrm{m}(x) &= g(x) \, g^{+}(x) \, \Delta(x),
		\label{eq:perturbation_decomposition_matched}
		\\ 
		\Delta_\mathrm{u}(x) &= g^{\perp}(x) \, g^{\perp +}(x) \, \Delta(x),
		\label{eq:perturbation_decomposition_unmatched}
	\end{align}
\end{subequations}
where $g^{+}(x)$ is the left pseudo-inverse of $g(x)$, i.e. $g^{+}(x) = (g^\top(x) \, g(x))^{-1} g^\top(x)$, and $g^{\perp}(x)$ is the full-rank annihilator of $g(x)$, i.e. a matrix with independent columns that span the null space of $g(x)$, i.e. $g^\top(x) \, g^{\perp}(x) = 0 \in \mathbb{R}^{1\times (n-1)}$ with $\mathrm{rank}(g^\perp) = n-1$. 
The following result establishes the restrictions, which \eqref{eq:model_uncertainty_without_unmatched_component} imposes on the model uncertainty $\Delta(\cdot)$ in original coordinates $x$ of the flat system \eqref{eq:system}.
\begin{thm}\label{thm:perturbation_in_original_coordinates}
	The uncertainty $\phi_\mathrm{u}\equiv 0$ in \eqref{eq:pertrubation_transformed_unmatched} if and only if $\Delta_\mathrm{u}(x) \equiv0$ in \eqref{eq:perturbation_decomposition}.
\end{thm}
\begin{proof}
	We first show that $\Delta_\mathrm{u}(x) \equiv 0$ is sufficient for \eqref{eq:model_uncertainty_without_unmatched_component}, and then establish the necessity of $\Delta_\mathrm{u}(x) \equiv 0$ by contradiction.\\
	$(\Leftarrow)$ 
	Let $\Delta_\mathrm{u}(x) = 0$.
	Evaluating \eqref{eq:perturbation_decomposition} yields $\Delta(x) = \Delta_\mathrm{m}(x) = g(x) \, \bar{\Delta}(x)$ with scalar function $\bar{\Delta}(x) = g^{+}(x) \, \Delta(x)$. 
	Thus,
	\begin{align}\label{eq:matched_perturbation_remains_matched}
		\mathcal{L}_{\Delta}\mathcal{L}_{f}^{k}h(x) 
		= \mathcal{L}_{g  \bar{\Delta}} \mathcal{L}_{f}^{k}h(x)
		= \mathcal{L}_{g}\mathcal{L}_{f}^{k}h(x) \, \bar{\Delta}(x) 
	\end{align} 
	for $k \leq n-2$, i.e. for $k=0,1,...,n-2$.
	Noting that $\mathcal{L}_{g}\mathcal{L}_{f}^{k}h(x) = 0$ by assumption in \eqref{eq:relative_degree_condition}, we have that \eqref{eq:model_uncertainty_without_unmatched_component} is satisfied.\\
	$(\Rightarrow)$
	Evaluating \eqref{eq:model_uncertainty_without_unmatched_component} with the decomposition \eqref{eq:perturbation_decomposition_sum} yields
	\begin{align*}
		\mathcal{L}_{\Delta_\mathrm{m} + \Delta_\mathrm{u}}  \mathcal{L}_{f}^{k}h(x)
		= \mathcal{L}_{\Delta_\mathrm{m}}  \mathcal{L}_{f}^{k}h(x) + \! \mathcal{L}_{\Delta_\mathrm{u}}  \mathcal{L}_{f}^{k}h(x)
		= 0
	\end{align*}
	for $k \leq n-2$.
	As discussed for \eqref{eq:matched_perturbation_remains_matched}, we have $ \mathcal{L}_{\Delta_\mathrm{m}}  \mathcal{L}_{f}^{k}h(x) = 0, k \leq n-2$ by construction of $\Delta_\mathrm{m}$ in \eqref{eq:perturbation_decomposition_matched}.
	Thus, 
	\begin{align}\label{eq:proof_unmatched_lie_derivatives_decompostion}
		\!
		\mathcal{L}_{\Delta_\mathrm{u}}  \mathcal{L}_{f}^{k}h(x) 
		\!=\! \textstyle ( \! \pdv{}{x}(
		\mathcal{L}_{f}^{k}h(x)
		) ) \Delta_\mathrm{u}(x) 
		\!=\! \Lambda_{k}(x)  \Delta_\mathrm{u}(x) 
		\!= \! 0
	\end{align}
	for all $k\leq n-2$, where we introduce the notation $\Lambda_k(x) \coloneqq \pdv{}{x}(\mathcal{L}_{f}^{k}h(x)) $ for brevity of the following expressions.
	
	For proof by contradiction, suppose that $\Delta_\mathrm{u}(x) \neq 0$.
	We show that $\Delta_\mathrm{u}(x)$ is not orthogonal to at least one of the non zero vectors $\Lambda_0^\top(x),...,\Lambda_{n-2}^\top(x)$.
	As discussed in \cite{Mar1995} and \cite{Kha2015}, the Jacobian $H(x) = \pdv{\tau(x)}{x}$ of the diffeomorphism $\tau(x)$ in \eqref{eq:state_transformation} is nonsingular.
	Thus, the rows of $H(x)$, which are given by $\Lambda_0(x),...,\Lambda_{n-2}(x),\Lambda_{n-1}(x) \neq 0$, are linearly independent.
	Moreover, with inequality \eqref{eq:relative_degree_condition_zero_before_input_channel} we have that $\mathcal{L}_{g}\mathcal{L}_{f}^{k}h(x) = \Lambda_{k}(x) g(x) =0$ for $k \leq n-2$, which is that the first $n-1$ rows $\Lambda_0(x),...,\Lambda_{n-2}(x)$ of $H(x)$ are orthogonal to $g^\top(x) \neq 0$.
	Hence, the $n$ linearly independent vectors $\Lambda_0^\top(x),...,\Lambda_{n-2}^\top(x),g(x)$ span $\mathbb{R}^n$, and with $g^\top(x) \, g^{\perp}(x) = 0$ it is readily verified that $\Delta_\mathrm{u}(x)$ in \eqref{eq:perturbation_decomposition_unmatched} is orthogonal to $g(x)$ by construction, i.e. $g^\top(x) \Delta_\mathrm{u}(x) = 0$.
	Noting that $\Delta_\mathrm{u}(x)$ cannot be orthogonal to all vectors of $\mathbb{R}^n$, we thus have that $\Delta_\mathrm{u}(x)$ is not orthogonal to at least one of the linearly independent vectors $\Lambda_0^\top(x),...,\Lambda_{n-2}^\top(x)$.
	Finally, $\Lambda_{k}(x) \, \Delta_\mathrm{u}(x) \neq 0$ for at least one $k \leq n-2$, which contradicts equality \eqref{eq:proof_unmatched_lie_derivatives_decompostion}.
\end{proof}
Theorem~\ref{thm:perturbation_in_original_coordinates} shows that we have a vanishing unmatched uncertainty $\phi_\mathrm{u}(x) \equiv 0$ in transformed coordinates in \eqref{eq:dynamics_system_transformed_coordinates} if and only if the model uncertainty $\Delta(\cdot)$ is matched in the original coordinates in \eqref{eq:system}, i.e. $\Delta_\mathrm{u}(x) \equiv0$.
Matched and unmatched uncertainties $\Delta_\mathrm{m}(\cdot)$ and $\Delta_\mathrm{u}(\cdot)$ in original coordinates~$x$, remain matched and unmatched uncertainties $\phi_\mathrm{m}(\cdot)$ and $\phi_{\mathrm{u}}(\cdot)$ in transformed coordinates $\xi$, respectively. 
I.e. unmatched uncertainties cannot be rendered matched by the transformation~\eqref{eq:state_transformation}.


\section{Unmatched model uncertainties within the model-following control architecture}\label{eq:mfc_with_perturbation}
In this section, we introduce the model-following control architecture and discuss the MFC design approach, which we considered in our preliminary studies \cite{WilWR2021}, \cite{WilWR2022}, \cite{TieWR2024CDCSMC}, \cite{TieWR2024a} and \cite{TieWR2024}, respectively.
\begin{figure}
	\centering
	\hspace{-3ex}
	\includegraphics[width=1\linewidth]{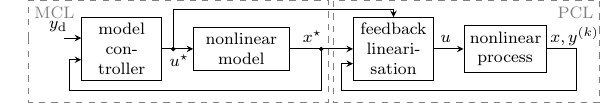}
	\caption{Blockdiagram of the model-following control (MFC) architecture with nonlinear process model.}
	\vspace{-2ex}
	\label{fig:MFC_blockdiagram_conventional}
\end{figure}
The model-following control scheme is shown in Figure \ref{fig:MFC_blockdiagram_conventional}.
The architecture consists of a model of the process simulated in the model control loop (MCL) and a process control loop (PCL) working on the actual process.
Given the nominal nonlinear model of the process
\begin{subequations}\label{eq:mfc_dynamics_mcl}
	\begin{align}
		\dot{x}^\star &= f(x^\star) + g(x^\star) \, u^\star,
		\\
		y^\star &= h(x^\star),
	\end{align}
\end{subequations}
with state $x^\star(t)\in \mathbb{R}^n$ and input $u^\star(t)\in \mathbb{R}$, the idea is to apply state feedback linearisation to both, the process \eqref{eq:system} and the model \eqref{eq:mfc_dynamics_mcl} to achieve asymptotic output tracking.
Transforming the dynamics of the open-loop system to Byrnes-Isidori form, we show that the MFC designs of our preliminary studies cannot be applied in presence of unmatched model uncertainties.
\subsection{State feedback linearisation of the MCL and the PCL}
For the MCL, we consider the state transformation
\begin{align}\label{eq:mfc_mcl_state_transformation}
	\xi^\star &= \tau(x^\star) = \begin{bmatrix}
		h(x^\star) & \! \mathcal{L}_{f}h(x^\star) & \! ... & \! \mathcal{L}_{f}^{n-1}h(x^\star)
	\end{bmatrix}^\top,
\end{align}
which transforms the dynamics of the nominal model \eqref{eq:mfc_dynamics_mcl} to Byrnes-Isidori form, i.e. 
\begin{align}
	\dot{\xi}^\star = A \, \xi^\star + B \big(
	\mathcal{L}_{f}^{n}h(x^\star) + \mathcal{L}_{g}\mathcal{L}_{f}^{n-1}h(x^\star) \, u^\star
	\big)
\end{align}
with output $y^\star = \xi_1^\star$, where the pair $(A,B)$ is in Brunovsk\'{y}-form.
Applying the state feedback linearisation control law 
\begin{align}\label{eq:u_star}
	u^\star = 
	\big(
	\mathcal{L}_{g}\mathcal{L}_{f}^{n-1}h(x^\star)	
	\big)^{-1}\big(
	- \mathcal{L}_{f}^{n}h(x^\star) 
	+ \omega^\star
	\big),
\end{align} 
with auxiliary control input $\omega^\star(t) \in \mathbb{R}^n$, we have 
\begin{align}\label{eq:system_linearised}
	\dot{\xi}^\star = A \, \xi^\star + B \, \omega^\star.
\end{align} 
For the PCL, we apply the feedback linearisation control law
\begin{align}\label{eq:u_mfc_pcl}
	u = 
	\big(
	\mathcal{L}_{g}\mathcal{L}_{f}^{n-1}h(x)
	\big)^{-1}\big(
	- \mathcal{L}_{f}^{n}h(x) 
	+ \omega^\star
	+ \tilde{\omega}
	\big),
\end{align}
with auxiliary control input $\tilde{\omega}(t) \in \mathbb{R}$ to the process \eqref{eq:system}.
Defining the error $\tilde{\xi} \coloneqq \xi - \xi^\star$, we obtain the dynamics of the PCL by substituting $u$ in \eqref{eq:u_mfc_pcl} into \eqref{eq:dynamics_system_transformed_coordinates} and subtracting the dynamics of the model \eqref{eq:system_linearised}.
We have
\begin{align}\label{eq:mfc_dynamics_pcl_transformed}
	\dot{\tilde{\xi}} = A \, \tilde{\xi} + B \big(
	\tilde{\omega}	+\phi_\mathrm{m}(x)
	\big) + \phi_{\mathrm{u}}(x)
\end{align} 
with $\phi_\mathrm{m}(\cdot)$ and $\phi_{\mathrm{u}}(\cdot)$ in \eqref{eq:pertrubation_transformed}.
Note that the unmatched uncertainty $\phi_{\mathrm{u}}(\cdot)$ of the process dynamics \eqref{eq:dynamics_system_transformed_coordinates} is also unmatched in the dynamics of the PCL \eqref{eq:mfc_dynamics_pcl_transformed}, which are given in terms of the deviation of the process state $\xi$ from the model state~$\xi^\star$.

\subsection{Trajectory tracking}
With the state transformation $\xi = \tau(x)$ for $\tau(\cdot)$ in \eqref{eq:state_transformation}, we achieve asymptotic output tracking $y \rightarrow y_\mathrm{d}$ by enforcing that the external state $\xi$ of the process tracks the desired external state
\begin{align}\label{eq:desired_state}
	\xi_\mathrm{d} \coloneqq \big[\begin{matrix}
		y_\mathrm{d} & \dot{y}_\mathrm{d} & ... & y_\mathrm{d}^{(n-1)}
	\end{matrix}\big]^\top.
\end{align}
Note that $\xi_{\mathrm{d}}$ is bounded by assumption.
For MFC trajectory tracking of $\xi_{\mathrm{d}}$, we typically apply a trajectory tracking control law $\omega^\star$ to the MCL \eqref{eq:system_linearised} and stabilise the dynamics of the PCL \eqref{eq:mfc_dynamics_pcl_transformed}, see e.g. \cite{TieWR2024a}.
Enforcing both, asymptotic tracking $\xi \rightarrow \xi_{\mathrm{d}}$ in the MCL and asymptotic convergence $\tilde{\xi} \rightarrow 0$ of the state of the PCL, we have convergence of the tracking error $	\xi - \xi_{\mathrm{d}} 
= \xi^\star + \tilde{\xi} - \xi_{\mathrm{d}}$.

With the dynamics of the MCL \eqref{eq:system_linearised} in Brunovsk\'{y}-form, we readily obtain a tracking control law, see e.g. \cite{TieWR2024a}.
However, for unmatched uncertainties $\phi_{\mathrm{u}}(x) \not \equiv 0$ the dynamics of the PCL \eqref{eq:mfc_dynamics_pcl_transformed} are not in Byrnes-Isidori form.
Using feedback $\tilde{\omega}$ of the state $\tilde{\xi}$, we cannot ensure stability of the PCL for non-vanishing unmatched uncertainty $\phi_\mathrm{u}(\cdot)$ in the transformed coordinates $\tilde{\xi}$.
The control designs of the PCL, which we consider in our studies \cite{WilWR2021}, \cite{WilWR2022}, \cite{TieWR2024CDCSMC}, \cite{TieWR2024a}, and \cite{TieWR2024}, cannot be applied in the presence of unmatched uncertainties.

\section{Main result}\label{sec:mfc_design}
\begin{figure}
	\centering
	\hspace{-3ex}
	\includegraphics[width=1\linewidth]{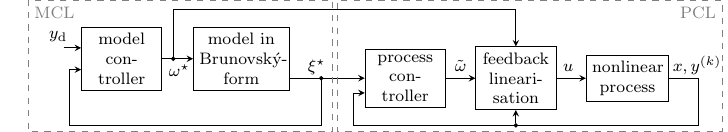}
	\caption{Blockdiagram of the model-following control (MFC) architecture with model of the linearised process dynamics in Brunovsk\'{y}-form.}
	\vspace{-2ex}
	\label{fig:MFC_blockdiagram_efficient}
\end{figure}
In this section we present our MFC design, which achieves trajectory tracking for unmatched model uncertainties.
We enforce trajectory tracking of $\xi_{\mathrm{d}}$ in the MCL, and stabilise the PCL by Lyapunov redesign using derivatives of the output~$y$, see~\eqref{eq:output_derivatives}.
We propose the implementation of the model-following control scheme as shown in Figure~\ref{fig:MFC_blockdiagram_efficient}
using the model \eqref{eq:system_linearised} in Brunovsk\'{y}-form.

\subsection{Control design}
We apply the control law  $u$ in \eqref{eq:u_mfc_pcl} with the following choice of the auxiliary inputs $\omega^\star$ and $\tilde{\omega}$.
For the MCL, we define the model tracking error
\begin{align}\label{eq:model_tracking_error}
	\tilde{\xi}^\star \coloneqq \xi^\star - \xi_{\mathrm{d}},
\end{align}
and consider the trajectory tracking law 
\begin{align}\label{eq:u_mfc_mcl}
	\omega^\star = y_\mathrm{d}^{(n)} + v(\xi^\star -\xi_{\mathrm{d}})
\end{align}
with error feedback $v(\xi^\star - \xi_{\mathrm{d}}) = v(\tilde{\xi}^\star)$ chosen such that the origin of the error dynamics
\begin{align}\label{eq:mcl_closed_loop_tracking_error_dynamics}
	\dot{\tilde{\xi}}^\star = A \, \tilde{\xi}^\star + B \, v(\tilde{\xi}^\star)
\end{align}
is asymptotically stable.

For the design of the PCL, we use the time-derivatives \eqref{eq:output_derivatives} of the output $y$ to define the state transformation
\begin{align*}
	\xi_{\mathrm{n}}
	&= \begin{bmatrix}
		y & \dot{y} & ... & y^{(n-1)}
	\end{bmatrix}^\top 
	= \tau_\mathrm{n}(x)
\end{align*}
with $\tau_\mathrm{n}(\cdot)$ in \eqref{eq:state_transformation_derivatives}, and introduce the error
\begin{align}\label{eq:state_pcl_derivatives}
	\tilde{\xi}_\mathrm{n} \coloneqq 
	\xi_{\mathrm{n}} - \xi^\star.
\end{align}
Let the state feedback $\tilde{v}(\tilde{\xi}_\mathrm{n})$ be chosen such that 
\begin{align*}
	\dot{\tilde{\xi}}_\mathrm{n} = A \, \tilde{\xi}_\mathrm{n} + B \, \tilde{v}(\tilde{\xi}_\mathrm{n})
\end{align*} 
is asymptotically stable with radially unbounded Lyapunov function $\tilde{V}(\cdot)$ such that
\begin{align}\label{eq:lyapunov_function_pcl}
	\pdv{\tilde{V}(\tilde{\xi}_\mathrm{n})}{\tilde{\xi}_\mathrm{n}} \big(
	A \, \tilde{\xi}_\mathrm{n} + B \, \tilde{v}(\tilde{\xi}_\mathrm{n})
	\big) \leq - \tilde{W}(\tilde{\xi}_\mathrm{n}),
\end{align}
where $\tilde{W}:\mathbb{R}^n \mapsto \mathbb{R}$ is a continuous positive definite function.
We apply the control law 
\begin{align}\label{eq:u_mfc_pcl_numeric_derivatives}
	\tilde{\omega} 
	= \tilde{v}(\xi_\mathrm{n} - \xi^\star) + \tilde{v}_\mathrm{L}
	= \tilde{v}(\tilde{\xi}_\mathrm{n}) + \tilde{v}_\mathrm{L}
\end{align}
with discontinuous control component
\begin{subequations}\label{eq:mfc_lyapunov_redesign_control_component_numerical_derivatives}
	\begin{align}
		&\tilde{v}_\mathrm{L}
		= - \tilde{\Gamma}_\mathrm{n}(x,\xi^\star) \ \mathrm{sgn}\big(
		\tilde{w}( \xi_{\mathrm{n}} -\xi^\star)
		\big),
		\\
		&\tilde{w}( \xi_{\mathrm{n}} -\xi^\star)
		=\tilde{w}(\tilde{\xi}_\mathrm{n}) 
		= \textstyle \pdv{\tilde{V}(\tilde{\xi}_\mathrm{n})}{\tilde{\xi}_\mathrm{n}}\, B
	\end{align}
\end{subequations}	
with suitably chose gain $\tilde{\Gamma}_\mathrm{n}:\mathbb{R}^n \times \mathbb{R}^n \mapsto \mathbb{R}^+$, which may depend on both, the state $x$ of the process and the state $\xi^\star$ of the model.
Substituting $\omega^\star$ in \eqref{eq:u_mfc_mcl} and $\tilde{\omega} $ in \eqref{eq:u_mfc_pcl_numeric_derivatives} into \eqref{eq:u_mfc_pcl}, we obtain the control law
\begin{align}\label{eq:u}
	u = \frac{
		- \mathcal{L}_{f}^{n}h(x) 
		+ y_\mathrm{d}^{(n)} + v(\tilde{\xi}^\star)
		+ \tilde{v}(\tilde{\xi}_\mathrm{n}) 
		+ \tilde{v}_\mathrm{L}
	}{\mathcal{L}_{g}\mathcal{L}_{f}^{n-1}h(x)}.
\end{align}

\subsection{Efficient implementation of the MCL}

The formulation of the control law \eqref{eq:u} allows for a simplified implementation of the MFC scheme.
The classical MFC scheme, shown in Figure~\ref{fig:MFC_blockdiagram_conventional}, obtains the dynamics \eqref{eq:system_linearised} of $\xi^\star$ by applying the feedback linearisation control law $u^\star$ in \eqref{eq:u_star} to the nonlinear model \eqref{eq:mfc_dynamics_mcl}.

Consider the control law in \eqref{eq:u} in terms of $\omega^\star$ and $\tilde{\omega}$ as in \eqref{eq:u_mfc_pcl} and depicted in Figure \ref{fig:MFC_blockdiagram_efficient}.
The auxiliary control $\omega^\star$ of the MCL in \eqref{eq:u_mfc_mcl} is given by feedback of the model tracking error $\tilde{\xi}^\star = \xi - \xi_{\mathrm{d}}$ in transformed coordinates, and $\tilde{\omega}$ in \eqref{eq:u_mfc_pcl_numeric_derivatives} is feedback of the error  $\tilde{\xi}_\mathrm{n} = \xi_\mathrm{n} - \xi^\star$ with gain $\tilde{\Gamma}$ in \eqref{eq:mfc_lyapunov_redesign_control_component_numerical_derivatives}, which depends on $x$ and $\xi^\star$. 
The control law \eqref{eq:u} thus requires, the state $x$ and the errors $\tilde{\xi}^\star$ and $\tilde{\xi}_\mathrm{n}$.
But in particular, we do not require the state $x^\star$ of the nonlinear model \eqref{eq:mfc_dynamics_mcl} in original coordinates.
Instead we only need to simulate the feedback-linearised system of \eqref{eq:mfc_dynamics_mcl} in the MCL, which consists of a simple integrator chain.

Both implementations yield mathematically equivalent control laws, however the proposed implementation in Figure~\ref{fig:MFC_blockdiagram_efficient} neither requires the simulation of the nonlinear model \eqref{eq:mfc_dynamics_mcl} nor its feedback linearising control, which avoids numeric difficulties and requires less computational effort.

Note that this simplified implementation can also be applied to MFC for systems with relative degree $r<n$.

\subsection{Stability and asymptotic tracking}
Define the auxiliary uncertainty terms
\begin{subequations}\label{eq:auxilary_perturbation_lie_derivative}
	\begin{align}
		\Delta_1(x) &\coloneqq	\mathcal{L}_{f + \Delta}^{n}h(x) - 	\mathcal{L}_{f}^{n}h(x),
		\label{eq:auxilary_perturbation_lie_derivative_state}
		\\
		\Delta_2(x)& \coloneqq	\mathcal{L}_{g}\mathcal{L}_{f + \Delta}^{n-1}h(x) - 	\mathcal{L}_{g}\mathcal{L}_{f}^{n-1}h(x),
		\label{eq:auxilary_perturbation_lie_derivative_input}
		\\
		\Delta_3(x) &\coloneqq 	\big(\mathcal{L}_{g}\mathcal{L}_{f}^{n-1}h(x)\big)^{-1} \, \Delta_2(x).
		\label{eq:auxilary_perturbation_lie_derivative_input_normalised}
	\end{align}
\end{subequations}
Assuming bounding conditions on $\Delta_1$ and $\Delta_3$, our main result establishes the tracking capability of the closed loop, which consists of the process \eqref{eq:system} and the controller \eqref{eq:u}.
\begin{thm}\label{thm:stability_mfc_numerical_derivatives}
	Consider the closed loop of system \eqref{eq:system} and controller \eqref{eq:u}.
	Let $\Delta_1(x)$ and $\Delta_3(x)$ be bounded by known functions $\delta_1,\delta_3: \mathbb{R}^n \mapsto \mathbb{R}^+$ such that
	\begin{align}\label{eq:bound_perturbation_numerical_derivatives_lyapunov_redesign}
		|\Delta_1(x)| \leq\delta_1(x)
		\quad \text{ and } \quad 
		|\Delta_3(x)| \leq \delta_3(x) < 1
	\end{align}
	for all $x\in \mathbb{R}^n$.
	Then, $\lim_{t \rightarrow \infty}(y(t)-y_\mathrm{d}(t)) = 0$ with bounded solution $x$ if the state-dependent gain $\tilde\Gamma_{\mathrm{n}}(x,\xi^\star)$ for all $x,\xi^\star \in \mathbb{R}^n$ satisfies
	\begin{align}\label{eq:mfc_numeric_derivative_lyapunov_redesign_stabilising_gain}
		\tilde{\Gamma}_\mathrm{n}(x,\xi^\star)
		\geq \frac{\delta_1(x) \!+\! \delta_3(x) \big|
			\!- \! \mathcal{L}_{f}^{n}h(x) 
			\!+\! y_\mathrm{d}^{(n)} 
			\! \!+\! v(\tilde{\xi}^\star)
			\!+\! \tilde{v}(\tilde{\xi}_\mathrm{n})
			\big| }{1- \delta_3(x)}.
	\end{align}
\end{thm}
\begin{proof}
	We first show that the bounded tracking error
	\begin{align*}
		\xi_{\mathrm{n}} - \xi_{\mathrm{d}} 
		= \big[\begin{matrix}
			y - y_\mathrm{d} & \dot{y} - \dot{y}_\mathrm{d} & ... & y^{(n-1)} - y_\mathrm{d}^{(n-1)}
		\end{matrix}\big]^\top
	\end{align*}
	asymptotically converges to zero, and thus $\lim_{t \rightarrow \infty}(y(t)-y_\mathrm{d}(t)) = 0$, and then establish the boundedness of~$x$.\\
	i) $\xi_{\mathrm{n}} - \xi_{\mathrm{d}}  \rightarrow 0$:
	With model tracking error $\tilde{\xi}^\star$ in \eqref{eq:model_tracking_error} and state $\tilde{\xi}_\mathrm{n}$ in \eqref{eq:state_pcl_derivatives}, the tracking error $\xi_{\mathrm{n}} - \xi_{\mathrm{d}}$ can be written as
	\begin{align}\label{eq:tracking_error}
		\xi_{\mathrm{n}} - \xi_{\mathrm{d}}
		= \xi^\star + \tilde{\xi}_{\mathrm{n}} - \xi_{\mathrm{d}}
		= \tilde{\xi}^\star + \tilde{\xi}_{\mathrm{n}}.
	\end{align}
	We show that both, $\tilde{\xi}^\star$ and $\tilde{\xi}_\mathrm{n}$ asymptotically converge to~zero.\\
	1) $\tilde{\xi}^\star \rightarrow 0$: 
	Note that the desired state $\xi_\mathrm{d}$ in \eqref{eq:desired_state} satisfies 
	\begin{align}\label{eq:desired_state_dynamics}
		\dot{\xi}_\mathrm{d} = A \, \xi_\mathrm{d} + B \, y_\mathrm{d}^{(n)}
	\end{align}
	by construction.
	We obtain the open-loop dynamics of $\tilde{\xi}^\star$ by subtracting \eqref{eq:desired_state_dynamics} from the dynamics \eqref{eq:system_linearised} of the model
	\begin{align*}
		\dot{\tilde{\xi}}^\star = A \, \tilde{\xi}^\star + B \big(
		\omega^\star - y_\mathrm{d}^{(n)}
		\big).
	\end{align*}
	Substituting the control law $\omega^\star$ from \eqref{eq:u_mfc_mcl} yields the closed-loop dynamics \eqref{eq:mcl_closed_loop_tracking_error_dynamics}, which are asymptotically stable by design.
	Thus, the solution $\tilde{\xi}^\star$ is bounded and $\lim_{t \rightarrow \infty}\tilde{\xi}^\star(t) = 0$.\\
	2) $\tilde{\xi}_\mathrm{n} \rightarrow 0$:
	The dynamics of $\xi_\mathrm{n} = \tau_\mathrm{n}(x)$ in \eqref{eq:state_transformation_derivatives} are given by
	\begin{align}\label{eq:dynamics_output_derivatives}
		\dot{\xi}_\mathrm{n} = A \, \xi_{\mathrm{n}} + B\big(
		\mathcal{L}_{f+\Delta}^{n}h(x) + \mathcal{L}_{g}\mathcal{L}_{f+\Delta}^{n-1}h(x) \, u
		\big).
	\end{align}
	Note that \eqref{eq:dynamics_output_derivatives} can be written as 
	\begin{align*}
		\dot{\xi}_\mathrm{n} \!=\! A \, \xi_{\mathrm{n}} \!+\! B\big(
		\mathcal{L}_{f}^{n}h(x) \!+\! \mathcal{L}_{g}\mathcal{L}_{f}^{n-1}h(x) \, u \!+\! \Delta_1(x) \!+\! \Delta_2(x) \, u
		\big)
	\end{align*}
	with uncertainties $\Delta_1(\cdot),\Delta_2(\cdot)$ as defined in \eqref{eq:auxilary_perturbation_lie_derivative}.
	Substituting $u$ from \eqref{eq:u_mfc_pcl} and subtracting \eqref{eq:system_linearised}, we obtain the closed-loop dynamics of $\tilde{\xi}_{\mathrm{n}}$
	\begin{align}\label{eq:mfc_pcl_dynamics_numeric_derivatives}
		\dot{\tilde{\xi}}_{\mathrm{n}} = A \, \tilde{\xi}_{\mathrm{n}} + B \big(
		\tilde{\omega} 
		+ \Delta_1(x) 
		+ \Delta_2(x) \, u
		\big),
	\end{align}
	where $\tilde{\omega} = \tilde{v}(\tilde{\xi}_\mathrm{n}) + \tilde{v}_\mathrm{L}$ in \eqref{eq:u_mfc_pcl_numeric_derivatives}.
	Along the solution of \eqref{eq:mfc_pcl_dynamics_numeric_derivatives}, the time-derivative of the Lyapunov function $\tilde{V}(\cdot)$ in \eqref{eq:lyapunov_function_pcl} is given by
	\begin{align*}
		\dot{\tilde{V}} \!=\! \textstyle \pdv{\tilde{V}(\tilde{\xi}_{\mathrm{n}})}{\tilde{\xi}_{\mathrm{n}}}\big(
		A \, \tilde{\xi}_{\mathrm{n}} \! + \!  B \, \tilde{v}(\tilde{\xi}_{\mathrm{n}})
		\big)
		\!+\! \tilde{w}(\tilde{\xi}_{\mathrm{n}})\big(
		\tilde{v}_\mathrm{L}
		\!+\! \Delta_1(x)
		\!+\! \Delta_2(x) \, u
		\big)
	\end{align*}
	with $\tilde{w}(\tilde{\xi}_\mathrm{n})$ as defined in \eqref{eq:mfc_lyapunov_redesign_control_component_numerical_derivatives}.
	Using \eqref{eq:lyapunov_function_pcl} and noting that $\tilde{w}(\tilde{\xi}_\mathrm{n}) \, \tilde{v}_\mathrm{L} = - \tilde{\Gamma}(x,\xi^\star) \, |\tilde{w}(\tilde{\xi}_\mathrm{n})|$ by construction, we obtain an estimate of $\dot{\tilde{V}}$ by
	\begin{align*}
		\dot{\tilde{V}} \leq -\tilde{W}(\tilde{\xi}_\mathrm{n}) - \big| \tilde{w}(\tilde{\xi}_\mathrm{n})\big| \, \big(
		\Gamma(x,\xi^\star) 
		- |\Delta_1(x)| 
		- |\Delta_2(x) \, u |
		\big).
	\end{align*} 
	Note that with $u$ in \eqref{eq:u_mfc_pcl} and $\Delta_3(\cdot)$ in \eqref{eq:auxilary_perturbation_lie_derivative_input_normalised}, we have 
	\begin{align*}
		\big|\Delta_2(x) \, u\big| 
		&= \big| \Delta_3(x)  \big( 
		- \mathcal{L}_{f}^{n}h(x) 
		+ \omega^\star
		+ \tilde{\omega} 
		\big) \big|
		\\
		&\leq \big| \Delta_3(x)\big|  \big(
		|\!- \!\mathcal{L}_{f}^{n}h(x) 
		\! + \omega^\star 
		\! + \tilde{v}(\tilde{\xi}_\mathrm{n})| + \Gamma(x,\xi^\star)
		\big),
	\end{align*}
	where we obtain the inequality by substituting $\tilde{\omega} = \tilde{v}(\tilde{\xi}_\mathrm{n}) + \tilde{v}_\mathrm{L}$ with $|\tilde{v}_\mathrm{L}| =\Gamma(x,\xi^\star)$.
	Thus,
	\begin{align*}
		\dot{\tilde{V}} \leq - \tilde{W}(\tilde{\xi}_\mathrm{n}) - \big|\tilde{w}(\tilde{\xi}_\mathrm{n})\big|\Big(
		(1 - |\Delta_3(x)|) \, \tilde{\Gamma}_\mathrm{n}(x,\xi^\star)
		- |\Delta_1(x)|
		- |\Delta_3(x)|  \
		\big|\!- \! \mathcal{L}_{f}^{n}h(x) + \omega^\star + \tilde{v}(\tilde{\xi}_\mathrm{n})\big|
		\Big).
	\end{align*}
	It is readily verified that $\dot{\tilde{V}}\leq - \tilde{W}(\tilde{\xi}_\mathrm{n})$ whenever
	\begin{align*}
		\tilde{\Gamma}_\mathrm{n}(x,\xi^\star) \geq \frac{
			|\Delta_1(x)|
			+ |\Delta_3(x)| \ \big|
			\!- \! \mathcal{L}_{f}^{n}h(x) + \omega^\star + \tilde{v}(\tilde{\xi}_\mathrm{n})
			\big|
		}{1 - |\Delta_3(x)|}.
	\end{align*} 
	With $\omega^\star$ in \eqref{eq:u_mfc_mcl} and $|\Delta_1(x)| \leq \delta_1(x)$, $|\Delta_3(x)| \leq \delta_3(x)<1$, the gain $\tilde{\Gamma}_\mathrm{n}(x,\xi^\star)$ satisfies the inequality if  \eqref{eq:mfc_numeric_derivative_lyapunov_redesign_stabilising_gain} is satisfied.
	Thus, the time-derivative of $\tilde{V}(\tilde{\xi}_\mathrm{n})$ is negative for all $\tilde{\xi}_\mathrm{n} \neq~0$, and with radially unboundedness of $\tilde{V}(\cdot)$ we obtain boundedness of the solution $\tilde{\xi}_\mathrm{n}$ and $\lim_{t \rightarrow \infty}\tilde{\xi}_\mathrm{n}(t)=0$.\\
	ii) Boundedness of $x$:
	The state $\xi_{\mathrm{n}} = \tau_\mathrm{n}(x)$ can be written as $\xi_\mathrm{n} = \xi_{\mathrm{d}} + \tilde{\xi}^\star + \tilde{\xi}_{\mathrm{n}}$, see  \eqref{eq:tracking_error}, and we have boundedness of $\xi_{\mathrm{d}}$ and $\tilde{\xi}^\star$, $\tilde{\xi}_{\mathrm{n}}$ by assumption and design, respectively.
	Thus, $\tau(x)$ \eqref{eq:tracking_error} is bounded.
	Noting that $\tau(\cdot)$ is a global diffeomorphism by assumption, we finally obtain boundedness of $x$.
\end{proof}
Theorem~\ref{thm:stability_mfc_numerical_derivatives} establishes that asymptotic tracking can be achieved for arbitrary initial states $x_0$ and $\xi_0^\star$ of the process~\eqref{eq:system} and the model \eqref{eq:system_linearised}, respectively, in presence of unmatched uncertainties satisfying the bounding condition \eqref{eq:bound_perturbation_numerical_derivatives_lyapunov_redesign}.
Similar to the analysis in \cite{WulPR2020}, the main idea of the proof is to show that the state transformation $\xi_{\mathrm{n}} = \tau_\mathrm{n}(x)$ with $\tau_\mathrm{n}(\cdot)$ in \eqref{eq:state_transformation_derivatives}, which considers the time-derivatives of the output $y$, renders all model uncertainties matched in \eqref{eq:dynamics_output_derivatives}. 
Introducing the auxiliary uncertainty terms $\Delta_1(\cdot)$ and $\Delta_2(\cdot)$ in \eqref{eq:auxilary_perturbation_lie_derivative}, we respectively separate the nominal terms $\mathcal{L}_{f}^{n}h(x)$ and $\mathcal{L}_{g}\mathcal{L}_{f}^{n}h(x)$ from the unknown nonlinearities $\mathcal{L}_{f+\Delta}^{n}h(x)$ and $\mathcal{L}_{g}\mathcal{L}_{f + \Delta}^{n-1}h(x)$, which contain the model uncertainty $\Delta(\cdot)$ in the input channel of \eqref{eq:dynamics_output_derivatives}.
The nominal control component $\tilde{v}$ of the control law $\tilde{\omega} $ in \eqref{eq:u_mfc_pcl_numeric_derivatives} stabilises the nominal dynamics, which we obtain for $\Delta(x) \equiv 0$.
The additive discontinuous component $\tilde{v}_\mathrm{L}$ in \eqref{eq:mfc_lyapunov_redesign_control_component_numerical_derivatives} of the Lyapunov redesign compensates the influence the matched uncertainties $\Delta_1(\cdot)$ and $\Delta_2(\cdot)$ on the time-derivative of the Lyapunov function $\tilde{V}(\cdot)$ for gain $\tilde{\Gamma}_\mathrm{n}(x,\xi^\star)$ satisfying \eqref{eq:mfc_numeric_derivative_lyapunov_redesign_stabilising_gain}.
With control coefficient $\mathcal{L}_{g}\mathcal{L}_{f + \Delta}^{n-1}h(x)$ depending on the model uncertainty $\Delta(\cdot)$, we require the gain $\tilde{\Gamma}_\mathrm{n}(\cdot,\cdot)$ to depend on the components $\omega^\star$ and $\tilde{v}$ of the control signal.
Thus, $\tilde{\Gamma}_\mathrm{n}(x,\xi^\star)$ depends on both, the state $x$ of the process and the state $\xi^\star$ of the model.
Together with bounding conditions of the form \eqref{eq:mfc_numeric_derivative_lyapunov_redesign_stabilising_gain}, dependency of the gain on components of the control signal is typical for discontinuous control design with unknown control coefficients, see e.g.
\cite[Chapter~2]{ShtEF2014}.

\subsection{Special case: matched model uncertainty}
Consider uncertainties $\Delta(x)$ in \eqref{eq:perturbation_decomposition} with only matched components, i.e. $\Delta(x) = \Delta_\mathrm{u}(x)$.
With \eqref{eq:relative_degree_condition} and \eqref{eq:model_uncertainty_without_unmatched_component} it is readily verified that 
\begin{align*}
	\mathcal{L}_{f + \Delta}^{k}h(x) = \mathcal{L}_{f}^{k}h(x),
	\quad k =0,1,...,n-1.
\end{align*}
Thus, the state transformations $\xi = \tau(x)$ with $\tau(\cdot)$ in \eqref{eq:state_transformation}, and $\xi_{\mathrm{n}} = \tau_\mathrm{n}(x)$ with $\tau_\mathrm{n}(\cdot)$ in \eqref{eq:state_transformation_derivatives} are equivalent.
For the auxiliary uncertainty terms in  \eqref{eq:auxilary_perturbation_lie_derivative} we then have
\begin{align*}
	\Delta_1(x) 
	&= \mathcal{L}_{\Delta} \mathcal{L}_{f}^{n-1}h(x) 
	= \phi_\mathrm{m}(x),
	\\ 
	\Delta_2(x) &= \Delta_3(x) = 0.
\end{align*}
Thus, the control components $\tilde{v}$ and $\tilde{v}_\mathrm{L}$ in \eqref{eq:u_mfc_pcl_numeric_derivatives} can be evaluated for $\tilde{\xi} = \xi - \xi^\star = \xi_{\mathrm{n}} - \xi^\star = \tilde{\xi}_\mathrm{n}$, respectively.
Moreover, with $\delta_3(x) = 0$ as bound for $|\Delta_3(x)|$, the condition~\eqref{eq:mfc_numeric_derivative_lyapunov_redesign_stabilising_gain} for asymptotic tracking simplifies to $\tilde{\Gamma}(x) \geq \delta_1(x)$.
Thus, the gain $\tilde{\Gamma}_\mathrm{n}$ is independent of the control signal $\omega^\star$.
The control law $u$ in \eqref{eq:u} then reads
\begin{align}\label{eq:u_pcl_matched_perturbation}
	\! u \!=\! \frac{
		\!-\! \mathcal{L}_{f}^{n}h(x) 
		\!+\!  y_\mathrm{d}^{(n)} \!
		\!+\! v(\tilde{\xi}^\star)
		\!+\! \tilde{v}(\tilde{\xi}) 
		\!-\! \tilde{\Gamma}_\mathrm{n}(x) \,  \mathrm{sgn}(\tilde{w}(\tilde{\xi}))
	}{\mathcal{L}_{g}\mathcal{L}_{f}^{n-1}h(x)},
\end{align}
where the gain $\tilde{\Gamma}_\mathrm{n}$ depends on the state $x$ of the process, only, and both, the feedback $\tilde{v}(\cdot)$ and the discontinuous term are evaluated for $\tilde{\xi}$. 

\subsection{Special case: initialising the model on the reference}
When initialising the process model \eqref{eq:system_linearised} on the reference $\xi_{\mathrm{d}}$, i.e. when selecting $\xi^\star(0) = \xi_{\mathrm{d}}(0)$, we obtain a trivial initial tracking error $\tilde{\xi}^\star(0) = \xi^\star(0) - \xi_{\mathrm{d}}(0) = 0$, and thus a trivial solution $\tilde{\xi}^\star  \equiv0$ of the asymptotically stable error dynamics \eqref{eq:mcl_closed_loop_tracking_error_dynamics}.
Thus, $\xi^\star \equiv \xi_{\mathrm{d}}$, i.e. the MCL replicates the desired external state with $\omega^\star = y_\mathrm{d}^{(n)} + v(0) = y_\mathrm{d}^{(n)}$ in \eqref{eq:u_mfc_mcl}.
For the PCL, the input $\tilde{\omega}$ in \eqref{eq:u_mfc_pcl_numeric_derivatives} is given for $\tilde{\xi}_{\mathrm{n}} = \xi_{\mathrm{n}} - \xi^\star = \xi_{\mathrm{n}} - \xi_{\mathrm{d}}$.
The control law in \eqref{eq:u} reduces to the single-loop control law
\begin{align}\label{eq:u_sl}
	u = \frac{
		- \mathcal{L}_{f}^{n}h(x) 
		\!+ \! y_\mathrm{d}^{(n)}
		\!+ \!  \tilde{v}(\bar{\xi}_{\mathrm{n}})
		\!-\! \tilde{\Gamma}_\mathrm{n}(x,\xi_{\mathrm{d}}) \, \mathrm{sgn}(\tilde{w}(\bar{\xi}_\mathrm{n}))
	}{\mathcal{L}_{g}\mathcal{L}_{f}^{n-1}h(x)},
\end{align}
with tracking error $\bar{\xi}_\mathrm{n} =  \xi_{\mathrm{n}} - \xi_{\mathrm{d}}$.
This shows that Theorem~\ref{thm:stability_mfc_numerical_derivatives} can be applied to establish the tracking capability of the single-loop control system resulting in the following corollary.

\begin{cor}\label{cor:stability_sl_numerical_derivatives}
	Consider the closed loop of system \eqref{eq:system} and controller \eqref{eq:u_sl}.
	Let $\Delta_1(x)$ and $\Delta_3(x)$ be bounded by known functions $\delta_1,\delta_3: \mathbb{R}^n \mapsto \mathbb{R}^+$ such that
	\begin{align*}
		|\Delta_1(x)| \leq\delta_1(x)
		\quad \text{ and } \quad 
		|\Delta_3(x)| \leq \delta_3(x) < 1
	\end{align*}
	for all $x\in \mathbb{R}^n$.
	Then, $\lim_{t \rightarrow \infty}(y(t)-y_\mathrm{d}(t)) = 0$ with bounded solution $x$ if the state-dependent gain $\tilde\Gamma_{\mathrm{n}}(x,\xi_\mathrm{d})$ for all $x,\xi_\mathrm{d} \in \mathbb{R}^n$ satisfies
	\begin{align*}
		\tilde{\Gamma}_\mathrm{n}(x,\xi_\mathrm{d})
		\geq \frac{\delta_1(x) \!+\! \delta_3(x) \big|
			\!- \! \mathcal{L}_{f}^{n}h(x) 
			\!+\! y_\mathrm{d}^{(n)}
			\!+\! \tilde{v}(\bar{\xi}_\mathrm{n})
			\big| }{1- \delta_3(x)}.
	\end{align*}
\end{cor}

\section{Illustrative example}\label{sec:example}
Similar to \cite{WulPR2020}, we consider a system in strict feedback form, namely the perturbed integrator chain
\begin{align}\label{eq:example_system}
	\dot{\xi} = A \, \xi + B\big(
	u + \Phi_\mathrm{m}(\xi)
	\big) + \Phi_\mathrm{u}(\xi)
\end{align}
with pair $(A,B)$ in Brunovsk\'{y}-form, state $\xi(t) = [\xi_1(t),...,\xi_n(t)]^\top\in \mathbb{R}^n$ and output $y(t) = \xi_1(t)$.
The unmatched model uncertainty is 
\begin{align*}
	\Phi_\mathrm{u}(\xi) = \begin{bmatrix}
		\alpha_1 \, \xi_2 & \alpha_2 \, \xi_3 & ... & \alpha_{n-1} \, \xi_n & 0
	\end{bmatrix}^\top
\end{align*} 
with constants $\alpha_1,...,\alpha_{n-1}\in(-1,1)$.
Let the matched uncertainty $\Phi_\mathrm{m}:\mathbb{R}^n \mapsto \mathbb{R}$ be bounded by a known function $\delta_1(x)$, i.e. $|\Phi_\mathrm{m}(\xi)| \leq \delta(\xi)$ for all $\xi\in \mathbb{R}^n$.

The auxiliary terms $\Delta_1(\cdot),\Delta_2(\cdot)$ and $\Delta_3(\cdot)$ are obtained by evaluating \eqref{eq:auxilary_perturbation_lie_derivative} for $f(x) = A \, x$ and $g(x) = B$ with $\xi = x$.
Alternatively, $\Delta_1(\cdot),\Delta_2(\cdot)$ and $\Delta_3(\cdot)$ can be obtained by calculating the $n$th time-derivative of the output $y$.
For the first $n-1$ time-derivatives, we have $y^{(k)} = p_k \, \xi_{k+1},k = 1,2,...,n-1$, where $p_k \coloneqq  \prod_{i = 1}^{k} (1+\alpha_i)$.
Thus, with $\dot{\xi}_n = u + \phi_{\mathrm{m}}(\xi)$, the $n$th time-derivative reads
\begin{align}
	y^{(n)} 
	= p_{n-1} \big( u +\phi_{\mathrm{m}}(\xi)\big)
	= \Delta_1(\xi) + \Delta_2 \, u + u ,
\end{align}
where $\Delta_1(\xi) = p_{n-1} \, \phi_{\mathrm{m}}(\xi) $ and $\Delta_2 = \Delta_3  = p_{n-1} - 1$.
Let
\begin{align}\label{eq:example_bound_auxilary_nonlinearity}
	1-\rho \leq p_{n-1} \leq 1+\rho
	\ \text{ for some } \ 
	\rho \in (0,1).
\end{align}
We have $|\Delta_3| = |p_{n-1}-1|\leq \rho$ and $|\Delta_1(\xi)| = |p_{n-1} \phi_{\mathrm{m}}(\xi)|\leq (1+\rho) \, \delta(\xi)$ with bound $\delta(\xi)$ of $|\phi_\mathrm{m}(\xi)|$.

\subsection{Control design}
Define the desired state $\xi_{\mathrm{d}}$ \eqref{eq:desired_state} and $\xi_{\mathrm{n}}= \tau_\mathrm{n}(x)$ with $\tau_\mathrm{n}(\cdot)$ in \eqref{eq:state_transformation_derivatives}, and consider the process model with state $\xi^\star(t) = [\xi_1^\star(t),...,\xi_n^\star(t)]^\top \in \mathbb{R}^n$ and output $y^\star(t) = \xi_1^\star(t)$, where $\dot{\xi}^\star = A \, \xi^\star + B \, \omega^\star$.
Let the gains $k,\tilde{k} \in \mathbb{R}^n$ be chosen such that the matrices $(A-B \, k^\top)$ and $(A - B \, \tilde{k}^\top)$ are Hurwitz.
With linear feedback $\tilde{v}(\tilde{\xi_{\mathrm{n}}})$ of $\tilde{\xi}_{\mathrm{n}} = \xi_{\mathrm{n}} - \xi^\star$, a Lyapunov function $\tilde{V}(\cdot)$ satisfying \eqref{eq:lyapunov_function_pcl} is given by $\tilde{V}(\tilde{\xi}_\mathrm{n}) = \tilde{\xi}_\mathrm{n}^\top \tilde{P} \, \tilde{\xi}_\mathrm{n}$ with $\tilde{P} = \tilde{P}^\top >0$ solution of the Lyapunov equation $(A - B \, \tilde{k}^\top)^\top \tilde{P} + \tilde{P} 	(A - B \, \tilde{k}^\top) = -I$.
With $\tilde{w}(\xi_{\mathrm{n}} - \xi^\star) = 2 \, (\xi_{\mathrm{n}} - \xi^\star)^\top \tilde{P} \, B$ in \eqref{eq:mfc_lyapunov_redesign_control_component_numerical_derivatives}, the control law \eqref{eq:u} is given by
\begin{align}\label{eq:example_u}
	u \!=\! \omega^\star 
	\!-\! \tilde{k}^\top \!(\xi_\mathrm{n} \! -\! \xi^\star)
	\!- \! \tilde{\Gamma}_\mathrm{n}(\xi,\xi^\star)  \mathrm{sgn}\big(
	2 \, (\xi_\mathrm{n} \!-\! \xi^\star)^\top \tilde{P} B
	\big),
\end{align}
with input $\omega^\star = y_\mathrm{d}^{(n)}-k^\top(\xi^\star - \xi_\mathrm{d})$ of the model.
Application of Theorem~\ref{thm:stability_mfc_numerical_derivatives} yields asymptotic tracking capability of the closed loop for the gain $\tilde{\Gamma}_\mathrm{n} = \tilde{\Gamma}_\mathrm{n}(\xi,\xi^\star)$ given by
\begin{align*}
	\tilde{\Gamma}_\mathrm{n}
	= (1-\rho)^{-1}\big(
	(1+\rho) \, \delta(\xi) + \rho \big(|u^\star| + |\tilde{k}^\top(\xi_\mathrm{n}-\xi^\star)|\big)
	\big)
\end{align*}
with the bound $\rho$ from \eqref{eq:example_bound_auxilary_nonlinearity}.

\subsection{Simulation of the closed loop}
Consider the case $n = 3$ with $\alpha_1 = 0.5$, $\alpha_2 = -0.5$, and matched uncertainty $\phi_{\mathrm{m}}(\xi) = \xi_1^2 + \xi_2^2 + \xi_3^2$.
Let $y_\mathrm{d}(t) = \sin(t)$, $\xi(0) = [1,0,0]^\top$, and $\xi^\star(0) = [0.5,0,0]^\top$.
Note that the initial state $\xi^\star(0)$ of the MCL does not match the initial state $\xi(0)$ of the plant. 
With $\alpha_1 = 0.5$ and $\alpha_2 = -0.5$ we have $p_2 = 0.75$.
The auxiliary uncertainty $\Delta_3$ is given by $\Delta_3 = p_{2} -1 = -0.25$.
Moreover, $|\phi_{\mathrm{m}}(\xi)| \leq \delta(\xi)$ with $\delta(\xi) = \Vert \xi \Vert_2^2$ and $1-\rho \leq p_{2} \leq 1+\rho$ for $\rho = 0.25$.
Selecting the gains $k=[1,3,3]^\top$ and $\tilde{k}=[64,48,12]^\top$ places the poles of the MCL and the PCL at $-1$ and $-4$, respectively.

\begin{figure}[htb]
	\centering
	\includegraphics[width=.7\linewidth]{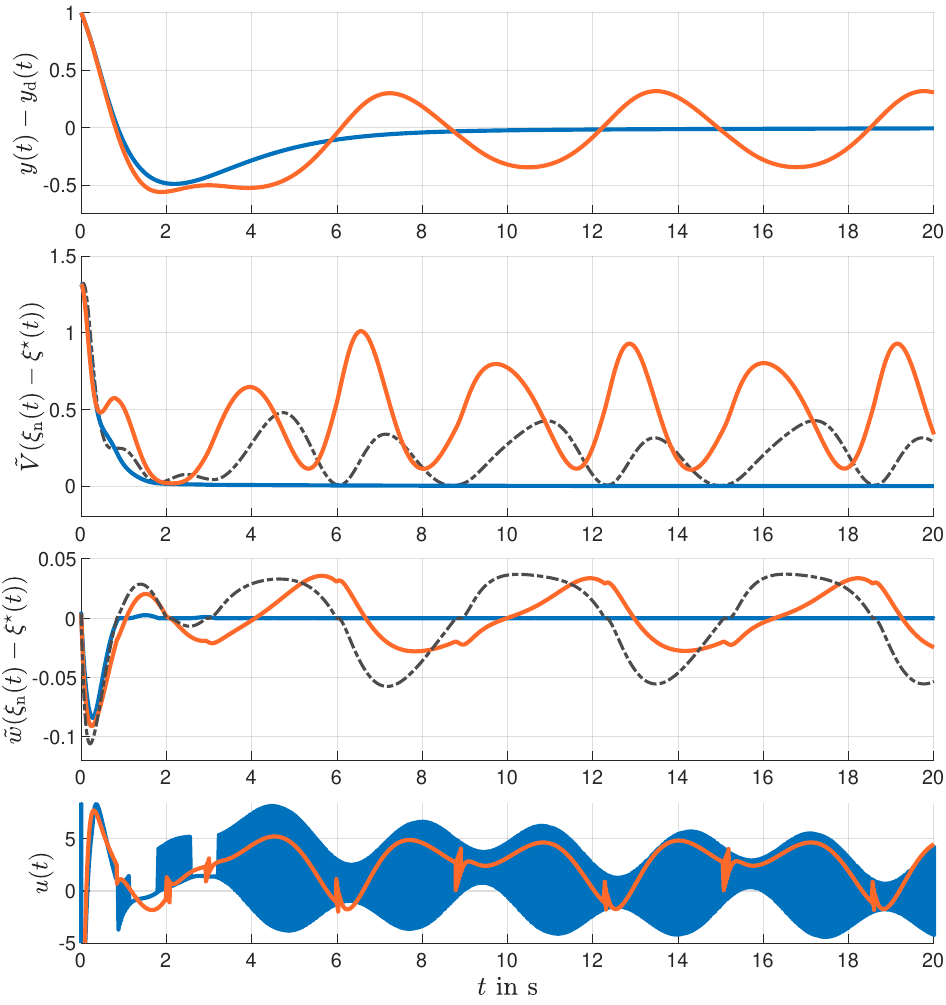}
	\caption{
		Simulation of the closed loop with  MFC \eqref{eq:example_u} in blue and the MFC \eqref{eq:u_pcl_matched_perturbation} in orange.
		Order from top to bottom: tracking error $y - y_\mathrm{d}$, Lyapunov function $\tilde{V}(\tilde{\xi}_\mathrm{n})$, auxiliary variable $\tilde{w}$, and control effort}
	\label{fig:example_tracking_error_time_domain}
\end{figure}

Figure shows the simulation results for the closed loop \eqref{eq:example_system}, \eqref{eq:example_u} in blue.
From top to bottom, the four plots show the tracking error $y - y_\mathrm{d}$, the value of the Lyapunov function $\tilde{V}$ evaluated along the solution $\tilde{\xi}_\mathrm{n} = \xi_{\mathrm{n}} - \xi^\star$, the auxiliary variable $\tilde{w}$ of the Lyapunov redesign evaluated along $\tilde{\xi}_\mathrm{n} $, and the control signal $u$.
Note that, by design the control signal continuous on the time intervals, where $\tilde{w}(\tilde{\xi}_\mathrm{n})$ is non zero.

For comparison, we consider the orange results, which are obtained for the closed loop consisting of the process \eqref{eq:example_system} and the controller \eqref{eq:u_pcl_matched_perturbation}, which compensates matched uncertainties only.
We obtain the control law by replacing the state $\xi_{\mathrm{n}}$ with $\xi$ in \eqref{eq:example_u}, and selecting the gain $\tilde{\Gamma}_\mathrm{n}(\xi) = \delta(\xi)$.
The dash-dotted line in the two middle plots of the Lyapunov function and the auxiliary variable is obtained by evaluating $\tilde{V}$ and $\tilde{w}$ along the solution $\tilde{\xi} = \xi - \xi^\star$ of the closed loop \eqref{eq:example_u}, \eqref{eq:u_pcl_matched_perturbation}, respectively.

The two middle plots show that the design \eqref{eq:example_u} enforces convergence for both, the auxiliary variable $\tilde{w}(\tilde{\xi}_\mathrm{n})$ and the Lyapunov function $\tilde{V}(\tilde{\xi}_\mathrm{n})$, and thus we have asymptotic output tracking \mbox{$\lim_{t \rightarrow \infty}(y(t)-y_\mathrm{d}(t)) = 0$} in presence of the unmatched uncertainty, with discontinuous control, as shown in the top and the bottom plot.
In comparison, the design in \eqref{eq:u_pcl_matched_perturbation} with feedback of the error $\tilde{\xi} = \xi -\xi^\star$ neither achieves convergence of $\tilde{w}(\tilde{\xi_{\mathrm{n}}})$ and $\tilde{V}(\tilde{\xi_{\mathrm{n}}})$ nor of $\tilde{w}(\tilde{\xi})$ and $\tilde{V}(\tilde{\xi})$ (dash-dotted lines), and the tracking error $y(t) - y_\mathrm{d}(t)$ does not vanish. 
By design the control signal \eqref{eq:u_pcl_matched_perturbation} is continuous on the time-intervals, where $\tilde{w}(\tilde{\xi})$ is non zero.
Thus, without convergence of $\tilde{w}(\tilde{\xi})$, the control signal \eqref{eq:u_pcl_matched_perturbation} is continuous for large parts of the simulation horizon.

\section{Conclusion}\label{sec:conclusion}
We show that unmatched uncertainties cannot be rendered matched by the state transformation required for the state feedback linearisation control. 
For the tracking problem for flat nonlinear systems we consider the MFC control scheme.
We propose a simple implementation for the MCL in the MFC that only requires the simulation of an integrator chain with linear feedback.
This reduces computational effort significantly and simplifies the implementation.
With Lyapunov redesign for feedback using the time-derivatives of the output, both, single-loop as well as MFC design can be robustified w.r.t. unmatched uncertainties.


\bibliographystyle{plain}
\bibliography{literatur.bib}

\begin{thebibliography}{10}

\bibitem{Azi2017}
Sajad Azizi.
\newblock Sufficient {{LMI}} conditions and {{Lyapunov}} redesign for the
  robust stability of a class of feedback linearized dynamical systems.
\newblock {\em ISA Transactions}, 68:90--98, 2017.

\bibitem{ByrI1991}
Christopher~I. Byrnes and Alberto Isidori.
\newblock Asymptotic stabilization of minimum phase nonlinear systems.
\newblock {\em IEEE Transactions on Automatic Control}, 36(10):1122--1137,
  1991.

\bibitem{CasF2006}
Fernando Casta{\~n}os and Leonid~M. Fridman.
\newblock Analysis and design of integral sliding manifolds for systems with
  unmatched perturbations.
\newblock {\em IEEE Transactions on Automatic Control}, 51(5):853--858, 2006.

\bibitem{CheC1998}
Tzuen-Lih Chern and Geeng-Kwei Chang.
\newblock Automatic voltage regulator design by modified discrete integral
  variable structure model following control.
\newblock {\em Automatica}, 34(12):1575--1585, 1998.

\bibitem{ChoW1995}
Yi-Shyong Chou and Wei Wu.
\newblock Robust controller design for uncertain nonlinear systems via feedback
  linearization.
\newblock {\em Chemical Engineering Science}, 50(9):1429--1439, 1995.

\bibitem{GuaS2001}
Guido~O. Guardabassi and Sergio~M. Savaresi.
\newblock Approximate linearization via feedback --- an overview.
\newblock {\em Automatica}, 37(1):1--15, 2001.

\bibitem{Gut1979}
Shaul Gutman.
\newblock Uncertain dynamical systems -- a {{Lyapunov}} min-max approach.
\newblock {\em IEEE Transactions on Automatic Control}, 24(3):437--443, 1979.

\bibitem{GutP1982}
Shaul Gutman and Zalman~J. Palmor.
\newblock Properties of min-max controllers in uncertain dynamical systems.
\newblock {\em SIAM Journal on Control and Optimization}, 20(6):850--861, 1982.

\bibitem{Isi1995}
Alberto Isidori.
\newblock {\em Nonlinear control systems}.
\newblock Springer, London, 3rd edition, 1995.

\bibitem{Kha2015}
Hassan~K. Khalil.
\newblock {\em Nonlinear {{Control}}}.
\newblock Pearson, Boston, 2015.

\bibitem{KhaS1982}
Hassan~K. Khalil and Ali Saberi.
\newblock Decentralized stabilization of nonlinear interconnected systems using
  high-gain feedback.
\newblock {\em IEEE Transactions on Automatic Control}, 27(1):265--268, 1982.

\bibitem{Lei1978}
George Leitmann.
\newblock Guaranteed ultimate boundedness for a class of uncertain linear
  dynamical systems.
\newblock {\em IEEE Transactions on Automatic Control}, 23(6):1109--1110, 1978.

\bibitem{Lev2009}
Arie Levant.
\newblock Non-homogeneous finite-time-convergent differentiator.
\newblock In {\em {{IEEE Conference}} on {{Decision}} and {{Control}}}, pages
  8399--8404, Shanghai, China, 2009.

\bibitem{LevLY2017}
Arie Levant, Miki Livne, and Xinghuo Yu.
\newblock Sliding-mode-based differentiation and its application.
\newblock In {\em {{IFAC World Congress}}}, pages 1699--1704, Toulouse, France,
  2017.

\bibitem{Mar1995}
Riccardo Marino and Patrizio Tomei.
\newblock {\em Nonlinear {{Control Design}}: {{Geometric}}, {{Adaptive}}, and
  {{Robust}}}.
\newblock Prentice Hall, 1995.

\bibitem{PosWR2020}
Tobias Posielek, Kai Wulff, and Johann Reger.
\newblock Analysis of sliding-mode control systems with unmatched disturbances
  altering the relative degree.
\newblock In {\em {{IFAC World Congress}}}, pages 5122--5128, 2020.

\bibitem{PosWR2023}
Tobias Posielek, Kai Wulff, and Johann Reger.
\newblock Analysis of sliding-mode control systems with relative degree
  altering perturbations.
\newblock {\em Automatica}, 148:5122--5128, 2023.

\bibitem{Pre1972}
G.~Preusche.
\newblock A two-level model following control system and its application to the
  power control of a steam-cooled fast reactor.
\newblock {\em Automatica}, 8(2):143--151, 1972.

\bibitem{RubEC2011}
Matteo Rubagotti, Antonio Estrada, Fernando Casta{\~n}os, Antonella Ferrara,
  and Leonid~M. Fridman.
\newblock Integral sliding mode control for nonlinear systems with matched and
  unmatched perturbations.
\newblock {\em IEEE Transactions on Automatic Control}, 56(11):2699--2704,
  2011.

\bibitem{SasK1988}
Shankar~S. Sastry and Petar~V. Kokotovi{\'c}.
\newblock Feedback linearization in the presence of uncertainties.
\newblock {\em International Journal of Adaptive Control and Signal
  Processing}, 2(4):327--346, 1988.

\bibitem{ShtEF2014}
Yuri Shtessel, Christopher Edwards, Leonid~M. Fridman, and Arie Levant.
\newblock {\em Sliding-mode control and observation}.
\newblock Birkh{\"a}user, 1st edition, 2014.

\bibitem{SugO1993}
Toshiharu Sugie and Koichi Osuka.
\newblock Robust model following control with prescribed accuracy for uncertain
  nonlinear systems.
\newblock {\em International Journal of Control}, 58(5):991--1009, 1993.

\bibitem{TieWR2024a}
Niclas Tietze, Kai Wulff, and Johann Reger.
\newblock Dynamic partial state-feedback revisited for output tracking using
  {{Lyapunov}} redesign and model-following control.
\newblock In {\em {{IEEE Conference}} on {{Decision}} and {{Control}}}, pages
  2882--2889, Milan, Italy, 2024.

\bibitem{TieWR2024CDCSMC}
Niclas Tietze, Kai Wulff, and Johann Reger.
\newblock Local stabilisation of nonlinear systems with time- and
  state-dependent perturbations using sliding-mode model-following control.
\newblock In {\em {{IEEE Conference}} on {{Decision}} and {{Control}}}, pages
  6620--6627, Milan, Italy, 2024.

\bibitem{TieWR2024}
Niclas Tietze, Kai Wulff, and Johann Reger.
\newblock A model-following control approach to peaking attenuation in
  high-gain partial state feedback for nonlinear systems.
\newblock In {\em {{IFAC Conference}} of {{Modelling}}, {{Identification}} and
  {{Control}} of nonlinear systems}, pages 7--12, Lyon, France, 2024.

\bibitem{WilWR2021}
Julian Willkomm, Kai Wulff, and Johann Reger.
\newblock Quantitative robustness analysis of model-following control for
  nonlinear systems subject to model uncertainties.
\newblock In {\em {{IFAC Conference}} on {{Modelling}}, {{Identification}} and
  {{Control}} of nonlinear {{Systems}}}, pages 167--172, Tokyo, Japan, 2021.

\bibitem{WilWR2022}
Julian Willkomm, Kai Wulff, and Johann Reger.
\newblock Set-point tracking for nonlinear systems subject to uncertainties
  using model-following control with a high-gain controller.
\newblock In {\em European {{Control Conference}}}, pages 1617--1622, London,
  United Kingdom, 2022.

\bibitem{WulPR2020}
Kai Wulff, Tobias Posielek, and Johann Reger.
\newblock Compensation of unmatched disturbances via sliding-mode control.
\newblock In Martin Steinberger, Martin Horn, and Leonid Fridman, editors, {\em
  Variable-{{Structure Systems}} and {{Sliding-Mode Control}}: {{From Theory}}
  to {{Practice}}}, Studies in {{Systems}}, {{Decision}} and {{Control}}, pages
  237--272. Springer International Publishing, Cham, 2020.

\bibitem{YouKU1977}
Kar-Keung~D. Young, Petar~V. Kokotovi{\'c}, and Vadim~I. Utkin.
\newblock A singular perturbation analysis of high-gain feedback systems.
\newblock {\em IEEE Transaction on Automatic Control}, 22(6):931--938, 1977.

\end{thebibliography}

\end{document}